\documentclass[aps,prb,twocolumn,superscriptaddress,showpacs,floatfix]{revtex4}
\usepackage{times}
\usepackage{graphicx}
\usepackage{epsfig}
\usepackage{amsmath}
\usepackage{amssymb}
\usepackage{natbib}
\usepackage{color}

\begin{document}

\title{The Magnetic Structure of Paramagnetic MnO}

\author{Joseph~A.~M.~Paddison}
\affiliation{Department of Chemistry, University of Oxford, Inorganic Chemistry Laboratory, South Parks Road, Oxford OX1 3QR, U.K.}
\affiliation{ISIS Facility, Rutherford Appleton Laboratory, Harwell Campus, Didcot OX11 0QX, U.K.}
\affiliation{School of Physics, Georgia Institute of Technology, 837 State Street, Atlanta, Georgia, 30332-0430, U.S.A.}

\author{Matthias~J.~Gutmann}
\affiliation{ISIS Facility, Rutherford Appleton Laboratory, Harwell Campus, Didcot OX11 0QX, U.K.}

\author{J.~Ross~Stewart}
\affiliation{ISIS Facility, Rutherford Appleton Laboratory, Harwell Campus, Didcot OX11 0QX, U.K.}

\author{Matthew~G.~Tucker}
\affiliation{ISIS Facility, Rutherford Appleton Laboratory, Harwell Campus, Didcot OX11 0QX, U.K.}
\affiliation{Diamond Light Source, Chilton, Oxfordshire, OX11 0DE, U.K.}
\affiliation{Spallation Neutron Source, Oak Ridge National Laboratory, Oak Ridge, Tennessee 37831, U.S.A.}

\author{Martin~T.~Dove}
\affiliation{School of Physics and Astronomy, Queen Mary University of London, Mile End Road, London E1 4NS, U.K.}

\author{David~A.~Keen}
\affiliation{ISIS Facility, Rutherford Appleton Laboratory, Harwell Campus, Didcot OX11 0QX, U.K.}

\author{Andrew~L.~Goodwin}
\email{andrew.goodwin@chem.ox.ac.uk}
\affiliation{Department of Chemistry, University of Oxford, Inorganic Chemistry Laboratory, South Parks Road, Oxford OX1 3QR, U.K.}
\date{\today}

\date{\today}

\begin{abstract}
Using a combination of single-crystal neutron scattering and reverse Monte Carlo refinements, we study the magnetic structure of paramagnetic MnO at a temperature (160\,K) substantially below the Curie-Weiss temperature $|\theta|\sim550$\,K. The microscopic picture we develop reveals a locally-ordered domain structure that persists over distances many times larger than the correlation length implied by direct analysis of the spin correlation function. Moreover, the directional dependence of paramagnetic spin correlations in paramagnetic MnO differs in some important respects from that of its incipient ordered antiferromagnetic state. Our results have implications for the understanding of paramagnetic states in weakly-frustrated systems, including high-temperature superconductors.

\end{abstract}

\maketitle

\section{Introduction}

In frustrated magnets, long-range magnetic order emerges at a temperature $T_{\rm c}$ substantially lower than the effective energy scale of magnetic interactions (\emph{i.e.}, the Curie-Weiss temperature $|\theta|$).\cite{Moessner_2006} A distinction is usually drawn between ``weak'' and ``strong'' frustration, associated with values of the frustration parameter $f=|\theta|/T_{\rm c}$ respectively smaller or larger than 10.\cite{Ramirez_1994} While the field has traditionally focussed on the exotic states accessible in strongly-frustrated systems,\cite{Bramwell_2011} weak frustration nevertheless plays a key role in the magnetic behaviour of a number of canonical antiferromagnets, including MnO.\cite{Anderson_1950} Of particular interest is the cooperative paramagnet (PM) regime $T_{\rm c}<T<|\theta|$ where magnetic interactions are still energetically relevant, yet incapable of driving long-range magnetic order. The ``fluctuating spin-stripe'' phases of cuprate superconductors are an ever-topical example of precisely such a state.\cite{Boothroyd_2011}

Given the importance of these canonical systems, it is perhaps surprising how little is known from an experimental viewpoint about the spin structures of cooperative PM states in weakly-frustrated magnets. The assumption is usually made that local magnetic order resembles that in the incipient ordered state,\cite{Renninger_1966} but is confined to small domains whose size is determined by the characteristic rate of decay of the spin correlation function.\cite{Kivelson_2003,Boothroyd_2011} This assumption is also implicit in conventional analysis of magnetic diffuse scattering \emph{via} Lorentzian fits.\cite{Zaliznyak_2015} In principle, the validity of this picture can be tested experimentally: magnetic diffuse scattering is sensitive to the three-dimensional spin correlations present in magnets---whether ordered or disordered.\cite{Blech_1964,Fennell_2014} When coupled with real-space refinement tools such as the reverse Monte Carlo (RMC) approach, this scattering can be used to generate experiment-driven atomic-scale models of the corresponding spin structure.\cite{Keen_1991,Mellergard_1998,McGreevy_2001,Tucker_2007,Paddison_2013,Paddison_2015}

In this study, we apply this combination of diffuse scattering and RMC analysis to determine the magnetic structure of MnO within its cooperative PM regime. Our analysis includes a new implementation of the RMC approach that allows direct fitting to single-crystal magnetic diffuse scattering. We find evidence of an extensive domain structure that is locally similar to the ordered AFM state but that also supports spin correlations forbidden by AFM order. Moreover, the domain sizes are substantially larger than suggested by direct analysis of the spin correlation function.

Our paper is arranged as follows. We begin with a short introduction to the magnetic behaviour of MnO. We then describe in turn the methods used in our study, including the new RMC implementation for single crystal diffuse scattering, and the results of our magnetic structure investigation of paramagnetic MnO. We conclude with a brief discussion of the implications of our results for other weakly-frustrated cooperative paramagnets.

\begin{figure}
\begin{center}
\includegraphics{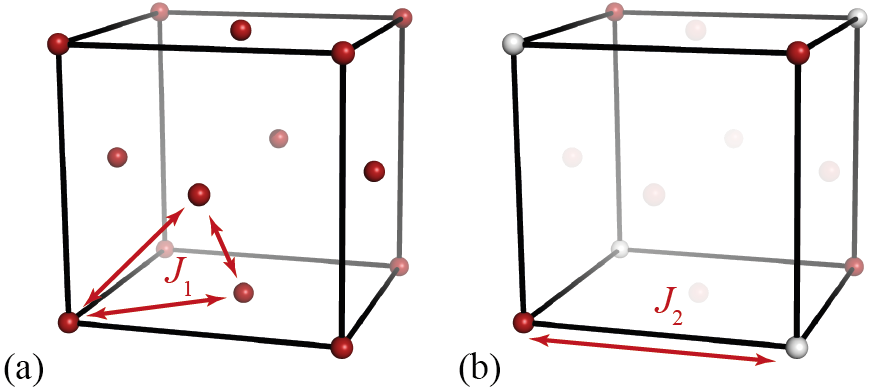}
\end{center}
\caption{\label{fig0}(a) Nearest-neighbour AFM interactions ($J_1$) are frustrated on the face-centred cubic lattice. (b) Next-nearest neighbour AFM interactions ($J_2$) are not frustrated and drive checkerboard ordering of the simple cubic sub-lattices.}\end{figure}

Above its magnetic ordering temperature $T_{\mathrm{N}}=118$\,K, MnO has the rock-salt structure, in which magnetic Mn$^{2+}$ ions ($S=5/2$, $L=0$) occupy a face-centred cubic lattice. The presence of weak frustration in MnO is indicated by a modest value of the frustration parameter $|\theta|/T_{{\rm {N}}}\approx5$.\cite{Tyler_1933}
which occurs because the frustrated antiferromagnetic (AFM) coupling between nearest neighbours is smaller than the unfrustrated next-nearest neighbour coupling [Fig.~\ref{fig0}].\cite{Pepy_1974} Below $T_{\mathrm{N}}$, long-range AFM order develops with magnetic propagation vector $\mathbf{k}=\left[\frac{1}{2}\frac{1}{2}\frac{1}{2}\right]^{*}$.\cite{Shull_1949} In the ordered AFM structure, spins are aligned parallel within $(111)$ planes, and the spin direction is reversed in adjacent $(111)$ planes.\cite{Roth_1958} The nearest-neighbour interactions within  $(111)$ planes are therefore frustrated, and a rhombohedral lattice distortion occurs in order to alleviate this frustration.\cite{Shaked_1988,Goodwin_2006}

Previous neutron-scattering studies have shown that structured magnetic
diffuse scattering is present above $T_{{\rm {N}}}$ (Refs.~\citenum{Hohlwein_2003,Blech_1964,Renninger_1966}) and short-range
spin correlations persist to $T\gtrsim1100$\,K.\cite{Mellergard_1998,Mellergard_1999} Yet, all previous measurements have been restricted to either individual reciprocal-space planes or the powder average, limiting the information content of the scattering pattern.\cite{Welberry_1998a} Advanced neutron-scattering instruments now allow measurement of essentially-complete three-dimensional (3D) diffuse-scattering patterns,\cite{Keen_2006,Bewley_2006} but a key problem remains: analysis of these very large datasets is usually computationally prohibitive.\cite{Welberry_1998a} Here, we develop an approach to allow rapid refinement of an atomic-scale model to magnetic diffuse-scattering datasets containing $>10^{6}$ data points. We demonstrate the success of this approach by fitting to the complete 3D magnetic diffuse-scattering pattern for MnO, allowing us to determine the relationship between PM and AFM structures.


\section{Methods}

\begin{figure}[b]
\begin{center}
\includegraphics{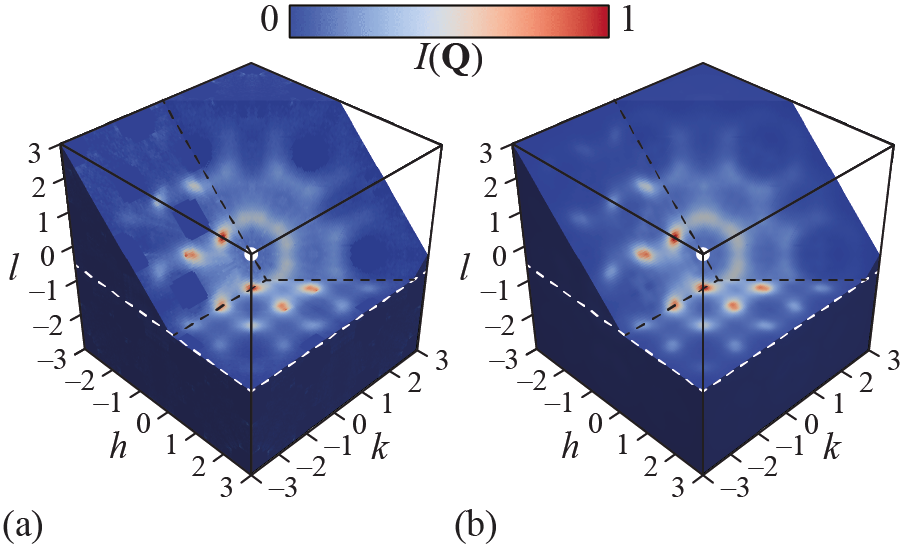}
\end{center}
\caption{\label{MnO_data_fit}(a) Experimental magnetic diffuse-scattering
data for paramagnetic MnO at $T=160$\,K. Nuclear Bragg peaks have been removed
from the data. (b) RMC fit to the experimental data shown in (a).
In (a) and (b), sections of the $(101)^\ast$, $(1\bar11)^\ast$, and $(001)^\ast$ reciprocal-space planes are shown. The $(001)^\ast$ plane is shifted by $-0.5$ reciprocal-lattice units along the $[001]^\ast$ direction in order to highlight the strongest diffuse scattering features; \emph{i.e.}, it is the $(h,k,-\frac{1}{2})^\ast$ plane. The centre of reciprocal space is indicated by a white circle.}\end{figure}

Single-crystal neutron-scattering data were collected at $T=160\,\mathrm{K}$ ($\simeq1.4T_{\mathrm{N}}$ and $0.3|\theta|$) using the SXD diffractometer at the ISIS neutron source.\cite{Keen_2006}  The data were corrected for instrumental background scattering by subtracting the scattering intensity from an empty sample
holder and were normalised using the incoherent scattering from a vanadium standard. The crystal structure (space group $Fm\bar{3}m$) was refined to the nuclear Bragg intensities using the \textsc{Jana} software package,\cite{Petricek_2014} using the lattice parameter $a = 4.4344(7)$\,\AA\ obtained from SXD at $T = 160$\,K. The data were binned in intervals of 0.04 reciprocal-lattice units, the $m\bar{3}m$ diffraction symmetry appropriate for MnO was applied, and nuclear Bragg peaks were removed by excising regions where the intensity exceeded a threshold value (plus a small surrounding volume). A 3D representation of the experimental data is shown in Fig.~\ref{MnO_data_fit}(a).

We employ reverse Monte Carlo (RMC) refinement \cite{McGreevy_1988,Tucker_2007,Paddison_2012} to fit spin configurations to our neutron-scattering data. In RMC refinement, a supercell of the crystallographic unit cell is generated and classical spin vectors are assigned to each site, whose orientations are refined to match experimental data. We use a cubic supercell of side length $R=12a$ ($N=6912$~spins) with periodic boundary conditions. Refinements are initialised with random spin orientations and are iterated to minimise a cost function
\begin{equation}
\chi^{2}=\sum_{\mathbf{Q}}\left[sI_{\mathrm{calc}}(\mathbf{Q})+B-I_{\mathrm{expt}}(\mathbf{Q})\right]^{2},
\end{equation}
where $I(\mathbf{Q})$ denotes the magnetic diffuse-scattering intensity at reciprocal-space position $\mathbf{Q}$, subscript ``calc" and ``expt" denote calculated and experimental data points, $s$ is a refined intensity scale factor, and $B$ is a refined flat-in-$\mathbf{Q}$ term which corrects for the significant incoherent scattering from Mn.\cite{Davis_1976} Results from four separate refinements were averaged to increase the statistical accuracy. The magnetic diffuse-scattering intensity is calculated as
\begin{equation}
I(\mathbf{Q}) \propto \left[f(Q)\right]^{2}\thinspace\exp(-U_{\mathrm{iso}}Q^{2})\sum_{\mathbf{G}}\left|\mathbf{F}(\mathbf{G})\right|^{2}W(\mathbf{Q}-\mathbf{G}),
\end{equation}
where $f(Q)$ is the Mn$^{2+}$ magnetic form factor,\cite{Brown_2004} $U_\mathrm{iso}=0.00509(9)$\,\AA$^{2}$ is the isotropic atomic displacement factor for Mn, and $\mathbf{G}$ is a reciprocal-lattice vector of the RMC supercell. The magnetic structure factor
\begin{equation}
\mathbf{F}(\mathbf{G}) = \sum_{i=1}^{N}\mathbf{S}_{i}^{\perp}\exp\left(\mathrm{i}\mathbf{G}\cdot\mathbf{r}_{i}\right),
\end{equation}
where $\mathbf{S}_{i}-\left[(\mathbf{S}_{i}\cdot\mathbf{G})\mathbf{G}\right]/G^{2}$ is the projection of the spin located at $\mathbf{r}_i$ perpendicular to $\mathbf{G}$. We use Lanczos resampling\cite{Duchon_1979} to interpolate values of $\left|\mathbf{F}(\mathbf{G})\right|^{2}$ at the experimentally-measured $\mathbf{Q}$-points by applying the weight function\cite{Duchon_1979} 
\begin{equation}
W(\mathbf{Q})=\prod_{\alpha}\mathrm{sinc}\left(Q_{\alpha}R/2\right)\mathrm{sinc}\left(Q_{\alpha}R/2m\right),
\end{equation}
where $\alpha \in \{x,y,z \}$ denotes Cartesian components, $m$ is an integer determining the interpolation accuracy, and $W(\mathbf{Q})\equiv 0$ outside the range $-m<Q_{\alpha}R/2\pi<m$. We take $m=4$, which allows the spin correlations to be calculated with $\pm 1\%$ accuracy for $0 \le r_{\alpha}\le 12$\,\AA. Importantly, the computational cost of updating $I(\mathbf{Q})$ after a single spin rotation scales approximately linearly with the number of $\mathbf{Q}$-points, and avoids redundant calculations necessary in current approaches where the supercell is divided into multiple ``sub-boxes".\cite{Butler_1992, Neder_2008} Our approach therefore allows rapid refinement of atomic-scale models to very large datasets (here, $\approx 1.5\times 10^6$ $\mathbf{Q}$-points).   



\section{Results}

The RMC fit to neutron-scattering data is shown in Fig.~\ref{MnO_data_fit}(b). Excellent agreement is achieved with the experimental data (the weighted-profile $R$-factor $R_{\mathrm{wp}}=8.3\%$). To the best of our knowledge, this
result represents the first time that an atomistic configuration has been
refined to a full 3D $I(\mathbf{Q})$ data set.

\begin{figure}
\begin{center}
\includegraphics{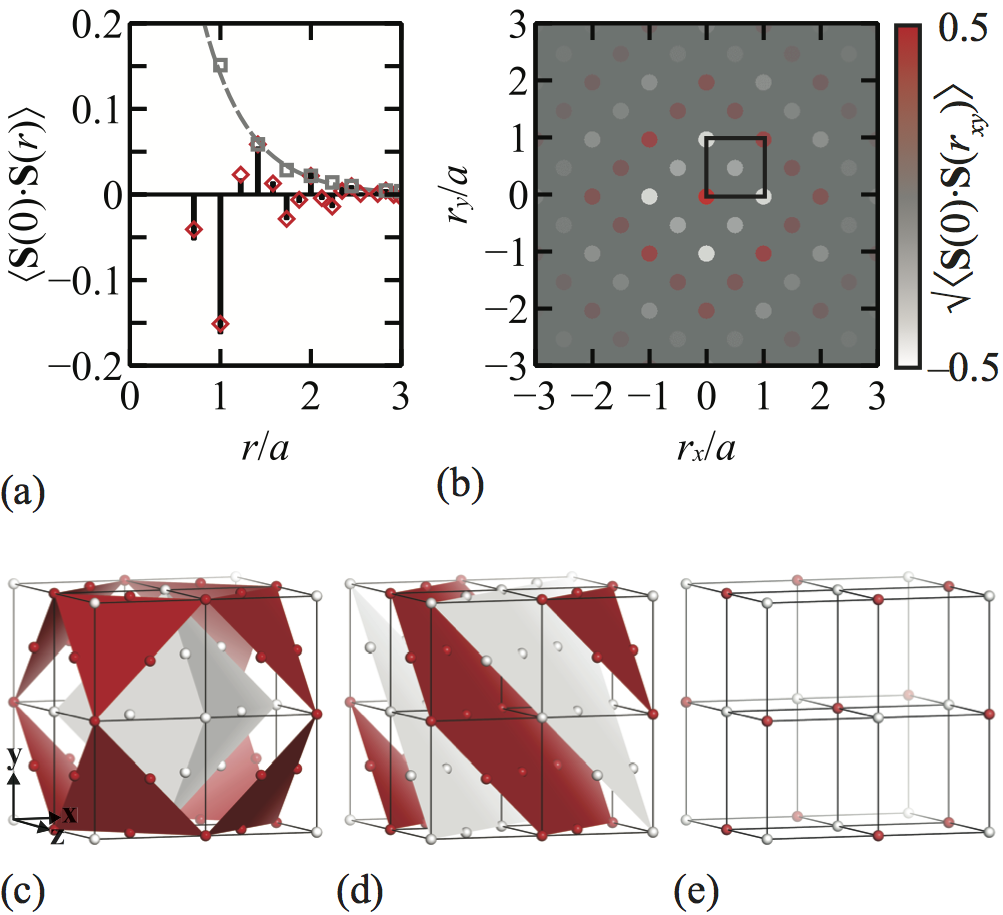}
\end{center}
\caption{\label{MnO_correlations} (a) Radial spin correlation function
$\langle \mathbf{S}(0)\cdot\mathbf{S}(r)\rangle $ for MnO
at $T=160\,$\,K. Black bars show results for the $J_1$-$J_2$ model described
in the text and red diamonds show results from RMC refinement to single-crystal
magnetic diffuse-scattering data. The dashed grey line shows the fit
of an exponential envelope to the RMC $|\langle \mathbf{S}(0)\cdot\mathbf{S}(r)\rangle|$,
which yields spin correlation length $\xi=2.258(1)$\,\AA. Grey squares show the $|\langle \mathbf{S}(0)\cdot\mathbf{S}(r)\rangle|$ values included in the fit. 
(b) 3D spin correlation function $\langle \mathbf{S}(\mathbf{0})\cdot\mathbf{S}(\mathbf{r})\rangle $
obtained from RMC refinement. The figure shows the $(xy0)$ plane (\emph{i.e.}, a cubic face). A square-root scale is use to show the longer-range correlations more clearly. The crystallographic unit cell is shown
as a black box. (c) Schematic representation of $\langle \mathbf{S}(\mathbf{0})\cdot\mathbf{S}(\mathbf{r})\rangle $
for MnO in the PM phase. Red areas indicate FM correlations and grey
areas AFM correlations. (d) Schematic representation of $\langle \mathbf{S}(\mathbf{0})\cdot\mathbf{S}(\mathbf{r})\rangle $
for a single domain of the ordered low-temperature AFM structure of
MnO. (e) Schematic representation of $\langle \mathbf{S}(\mathbf{0})\cdot\mathbf{S}(\mathbf{r})\rangle $
obtained for the AFM structure with the point symmetry of the Mn site in the PM state applied.}\end{figure}

The spin Hamiltonian of MnO has previously been characterised using inelastic neutron-scattering measurements in the ordered AFM
phase\cite{Pepy_1974} and diffuse-scattering measurements
of the $(110)^{*}$ plane in the PM phase.\cite{Hohlwein_2003} A Heisenberg model with AFM nearest and next-nearest neighbour exchange constants $J_{1}=-3.3\,\textrm{K}$ and $J_{2}=-4.6\,\textrm{K}$ provides a good description of the diffuse-scattering data at $T\approx160\,\textrm{K}$.\cite{Hohlwein_2003} As a check on our RMC refinement, we simulated this $J_{1}$-$J_{2}$ model at $T=160\,\textrm{K}$ using a direct Monte Carlo approach. Fig.~\ref{MnO_correlations}(a) compares
the radial spin correlation function $\langle \mathbf{S}(0)\cdot\mathbf{S}(r)\rangle $
obtained from RMC refinement with the results for the $J_{1}$-$J_{2}$ model. The trend in the correlations is identical between the two calculations; quantitatively, the difference in magnitude of the next-nearest neighbour correlation value is 7\%. The spin correlation length $\xi=2.258(1)$\,\AA\ $\simeq a/2$ was obtained by fitting $\exp(-r/\xi)$ to $|\langle\mathbf S(0)\cdot\mathbf S(r)\rangle|$ over the set of distances for which $|\langle\mathbf S(0)\cdot\mathbf S(r)\rangle|$ is larger than at all longer distances. We will come to show that local magnetic order persists over a length-scale substantially larger than $\xi$. Motivated by the evidence from $\gamma$-ray diffraction for a non-spherical distortion of the $d$-electron density in the PM phase,\cite{Jauch_2003} we also calculated the distribution of spin orientations from our RMC refinements but observed no statistically-significant anisotropy in the spin orientations. This result is consistent with the observation that the magnetic dipolar interaction is mainly responsible for magnetic anisotropy in MnO,\cite{Kaplan_1954,Keffer_1957} but its strength $DS(S+1)\approx 11$\,K (Ref.~\citenum{Pepy_1974}) is much smaller than the thermal energy at $T=160$\,K. The results from RMC refinement therefore agree closely with the $J_{1}$-$J_{2}$ Heisenberg model of paramagnetic MnO, validating the methodology of 3D RMC refinement.

Access to 3D spin configurations allows us to probe magnetic structure in more depth than given by radial spin correlation functions alone. Our particular interest is in understanding the relationship between the PM and AFM states in MnO. The $\langle \mathbf{S}(0)\cdot\mathbf{S}(r)\rangle $
function shown in Fig.~\ref{MnO_correlations}(a) already hints that the
PM correlations do not simply resemble the AFM correlations multiplied
by a decreasing function of distance. As expected from the relative
magnitudes of $J_{1}$ and $J_{2}$, the strongest correlation is
between next-nearest neighbours, for which AFM interactions are not
frustrated. However, significant AFM correlation is present at the
nearest-neighbour distance in the PM phase, whereas this correlation
is exactly zero for the ordered AFM state. This result implies that
the absence of long-range order allows frustrated nearest-neighbour
interactions to be partially satisfied in the PM phase. In order to
assess the influence of the frustrated geometry on the spin correlations,
we consider the 3D spin correlation function $\langle\mathbf{S}(\mathbf{0})\cdot\mathbf{S}(\mathbf{r})\rangle$. This function reveals the dependence of spin correlations on the lattice geometry, which is expected to be key in frustrated systems.\cite{Paddison_2013a} Fig.~\ref{MnO_correlations}(b)
shows that a distinctive pattern---hidden in the radial correlation
function---emerges in $\langle \mathbf{S}(\mathbf{0})\cdot\mathbf{S}(\mathbf{r})\rangle $.
The $\langle \mathbf{S}(\mathbf{0})\cdot\mathbf{S}(\mathbf{r})\rangle $
can be described as a set of nested octahedral shells, with the sign
of the spin correlations alternating between FM and AFM for successive
shells as distance is increased [Fig.~\ref{MnO_correlations}(c)].
As anticipated, this pattern extends over length-scales much greater than $\xi$. Hence, taking
each Mn atom in turn as the origin, Mn neighbours at coordinates $\mathbf{r}/a=\left[x,y,z\right]$
are (on average) ferromagnetically correlated if $x+y+z$ is even, and antiferromagnetically
correlated if $x+y+z$ is odd. The sign of $\langle \mathbf{S}(\mathbf{0})\cdot\mathbf{S}(\mathbf{r})\rangle $
is consistent with the fact that $J_{1}$ and $J_{2}$ interactions
are both AFM and the smallest number of exchange pathways which connects
two Mn\textsuperscript{2+} ions is given by $x+y+z$.

To what extent are these local correlations related to the spin structure of the low-temperature AFM phase? The description of the PM spin structure of MnO as a set of nested
octahedral shells [Fig.~\ref{MnO_correlations}(c)] is compared in Figs.~\ref{MnO_correlations}(d) and \ref{MnO_correlations}(e)
with, respectively, a single domain of the AFM structure and this same structure with the point symmetry of the Mn site in PM MnO ($m\bar3m$) applied. In a single domain of the AFM structure [Fig.~\ref{MnO_correlations}(d)],
spins are ferromagnetically aligned within $(111)$ planes and the
direction of spin alignment reverses between adjacent planes;\cite{Roth_1958}
hence, the PM correlations resemble the AFM structure viewed along
the $[111]$ direction. However, a sum over symmetry-equivalent AFM domain orientations [Fig.~\ref{MnO_correlations}(e)]
cannot fully describe the PM correlations, because $\langle \mathbf{S}(\mathbf{0})\cdot\mathbf{S}(\mathbf{r})\rangle $
must vanish for nearest-neighbour spins in this case. The RMC results [Fig.~\ref{MnO_correlations}(b)] are intermediate between Fig.~\ref{MnO_correlations}(c) and Fig.~\ref{MnO_correlations}(e): the signs of the spin correlations are described by Fig.~\ref{MnO_correlations}(c) but the magnitudes of the spin correlations are largest at the positions shown in Fig.~\ref{MnO_correlations}(e). Consequently, an interpretation of the PM phase in terms of local AFM order explains the strongest spin correlations but is nevertheless an oversimplification because the nature of the nearest-neighbour correlations is different in PM and AFM states. These results are entirely consistent with (i) the early theoretical studies of Refs.~\citenum{Lines_1965,Lines_1965b} based on the random-phase Green's function approximation and (ii) recent magnetic pair distribution function (mPDF) analysis of powder neutron scattering data.\cite{Frandsen_2015}

\begin{figure}[b]
\begin{center}
\includegraphics{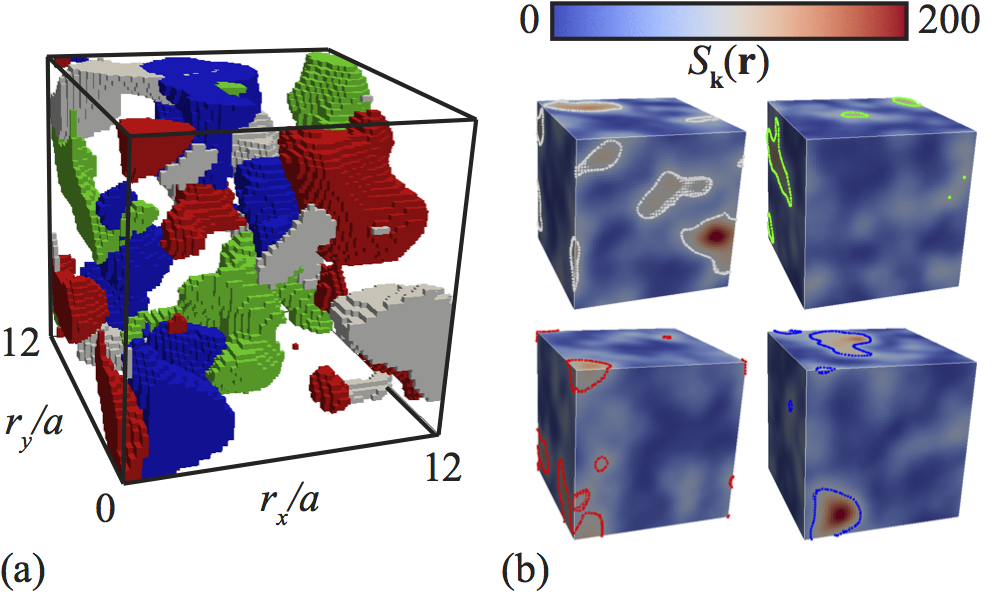}
\end{center}
\caption{\label{MnO_domains}Magnetic domain structure of MnO at $T=160$\,K
obtained as described in the text. In (a), different domains with
local periodicity $\mathbf{k}\in\langle \frac{1}{2}\frac{1}{2}\frac{1}{2}\rangle ^{*}$
are shown in different colours. The coloured regions show the threshold
where $S_{\mathbf{k}}(\mathbf{r})=85$ (the range of $S_{\mathbf{k}}(\mathbf{r})$
is between approximately zero and 200). In (b), the four images show
different $\mathbf{k}\in\langle \frac{1}{2}\frac{1}{2}\frac{1}{2}\rangle ^{*}$. The value of $S_{\mathbf{k}}(\mathbf{r})$ in each image is
shown using a blue-to-red colourmap. Coloured points indicate
the threshold where $S_{\mathbf{k}}(\mathbf{r})=85$, using the same colours  as (a). All images are shown in the same orientation.}\end{figure}

We proceed to explore the length-scale over which this modified AFM-like local order persists. In the conventional interpretation,
local AFM order is characterised by one of the four symmetry-equivalent $\mathbf{k}\in\langle \frac{1}{2}\frac{1}{2}\frac{1}{2}\rangle^{*}$.\cite{Renninger_1966,Kadanoff_1967}
It is assumed that different $\mathbf{k}$ are selected within separate regions of the crystal, so
that the overall cubic symmetry of the PM phase is preserved. To look for
such domain structure in the RMC spin configurations, we calculate
a local version of the magnetic scattering factor, 
\[
S_{\mathbf{k}}(\mathbf{r})=\left|\sum_{i}\mathbf{S}_{i}\exp\left(\frac{2\pi\mathrm{i}}{a}\mathbf{k}\cdot\mathbf{r}_{i}\right)\exp\left(-\frac{\left|\mathbf{r}-\mathbf{r}_{i}\right|}{2\xi}\right)\right|^{2},
\]
where $\mathbf{k}\in\langle \frac{1}{2}\frac{1}{2}\frac{1}{2}\rangle ^{*}$, $\mathbf{r}_{i}$ is the position
of spin $\mathbf{S}_{i}$ within the configuration, and the continuous variable $\mathbf{r}$
denotes position within the configuration.
The quantity $S_{\mathbf{k}}(\mathbf{r})$ is sensitive to local AFM order with modulation vector $\mathbf k$ at position $\mathbf{r}$. Fig.~\ref{MnO_domains}
shows that a representative RMC spin configuration contains many continuous regions within
which a single $\mathbf{k}$ dominates. Similar behaviour (not shown) is also observed for the $J_1$-$J_2$ model. These regions persist over a length-scale that is an order of magnitude larger than the value of $\xi$ determined by direct analysis of the spin correlation function and are largely non-overlapping, suggesting that a meaningful domain structure is indeed present in the PM phase of MnO.

\section{Concluding Remarks}

Our study has provided key experimental insight into the nature of the cooperative PM phase of MnO. We find a modified AFM-like local order that persists over continuous regions, each associated with one of four symmetry-equivalent modulation vectors $\mathbf{k}\in\langle \frac{1}{2}\frac{1}{2}\frac{1}{2}\rangle ^{*}$ and each spanning many unit cells.  The presence of local $\langle \frac{1}{2}\frac{1}{2}\frac{1}{2}\rangle^{*}$ periodicity provides a natural explanation for the experimental observation of dispersive (spin-wave-like) excitations in this phase.\cite{Betsuyaku_1978} The clearest difference between local spin correlations in the PM and AFM states is the relief of nearest-neighbour frustration in the former. This effect is analogous to the situation observed in some metallic glasses, where structural units can possess local icosahedral symmetry inconsistent with the long-range structural periodicity of crystalline arrangements.\cite{Luo_2004,Sheng_2006} The observation of non-trivial PM correlations in a weakly-frustrated system has implications for the interpretation of spin disorder in, \emph{e.g.}, ``fluctuating spin-stripe'' phases of high-temperature superconductors, which have traditionally been assumed to resemble ordered states over short length-scales.\cite{Kivelson_2003}

As a final point, we note that a separate result of our study is the development of an approach for refining spin configurations against the full 3D neutron-scattering pattern measurable using instruments such as SXD at ISIS ($>10^6$ $\mathbf{Q}$-points). This provides a practical model-independent alternative to traditional methods for interpreting diffuse-scattering data.\cite{Fennell_2012,Manuel_2009}


\section*{Acknowledgements}

J.A.M.P. and A.L.G. gratefully acknowledge financial support from the STFC, EPSRC (Grant EP/G004528/2) and ERC (Grant 279705). We are grateful to M. J. Cliffe (Cambridge) for helpful discussions.

\bibliography{jamp_full_refs}
\bibliographystyle{Science}
\end{document}